\documentclass{article}

\usepackage{amsmath,amssymb}

\def\X{{\cal X}}
\def\Y{{\cal Y}}
\def\Z{{\cal Z}}

\font\tenscr=rsfs10 scaled1100
\font\sevenscr=rsfs7 
\font\fivescr=rsfs5 
\skewchar\tenscr='177 
\skewchar\sevenscr='177
\skewchar\fivescr='177 
\newfam\scrfam 
\textfont\scrfam=\tenscr
\scriptfont\scrfam=\sevenscr 
\scriptscriptfont\scrfam=\fivescr

\def\linebreak{\hfill\break}

\def\bra<#1|{\langle #1\rvert}
\def\ket|#1>{\lvert#1 \rangle}
\def\braket<#1|#2>{\langle #1|#2 \rangle}

\def\bigpare(#1){\left(#1\right)}

\def\bigbra[#1]{\left[ #1 \right]}

\def\therefore{\mbox{\setbox0=\hbox{X}\hbox{$\ldotp$}\raise0.7\ht0\hbox{$\ldotp$}\hbox{$\ldotp$}} \quad }
\def\because{\mbox{\setbox0=\hbox{X}\raise0.7\ht0\hbox{$\ldotp$}\hbox{$\ldotp$}\raise0.7\ht0\hbox{$\ldotp$}}\kern0pt }

\def\bm#1{\boldsymbol{#1}}

\def\Frac(#1/#2){\left(\frac{#1}{#2}\right)}

\def\Eq#1{\begin{equation} #1 \end{equation}}

\def\Eqr#1{\begin{eqnarray} #1 \end{eqnarray}}

\def\Bitm{\begin{itemize}}
\def\Eitm{\end{itemize}}
\def\Blist#1#2{\begin{list}{#1}{\parsep=0pt \itemsep=0pt%
  \listparindent=0pt #2}}
\def\Elist{\end{list}}
\long\def\ignore#1#2{\def\ignoreflag{#1}\long\def\tmptext{#2}
  \ifnum\ignoreflag>1 #2 \fi}

\newcommand{\nn}{\nonumber}
\newcommand{\pd}{\partial}

\def\Xsp{{\rm X}}

\def\Ysp{{\rm Y}}
\def\X5sp{{\rm X}_5}
\def\Y3sp{{\rm Y}_3}
\def\Z3sp{{\rm Z}_3}

\numberwithin{equation}{section}

\def\abstract#1{\vskip 7mm 
        \begin{center}{\large Abstract}\par \smallskip
                \begin{minipage}[c]{12cm}
                        \small #1
                \end{minipage}
        \end{center}
}
\def\title#1{\begin{center}{\Large\bf #1}\end{center}}
\def\author#1{\vskip 5mm \begin{center}{#1}\end{center}}
\def\address#1{\begin{center}{\it #1}\end{center}}
\makeatletter
\@ifundefined{lesssim}{}{}
\@ifundefined{gtrsim}{}{}
\def\vereq#1#2{\lower3pt\vbox{\baselineskip1.5pt \lineskip1.5pt
\ialign{$\m@th#1\hfill##\hfil$\crcr#2\crcr\sim\crcr}}}
\makeatother

\begin{document}

\title{%
  Dynamical solution of supergravity
}
\author{%
  Pierre Binetruy\footnote{E-mail:Pierre.Binetruy@apc.univ-paris7.fr}, 
  Misao Sasaki\footnote{E-mail:misao@yukawa.kyoto-u.ac.jp}, 
  Kunihito Uzawa\footnote{E-mail:uzawa@ksc.kwansei.ac.jp}
}
\address{%
  $^1$Astroparticule et Cosmologie (UMR 7164), 
  Universite Paris 7, F-75205 Paris Cedex 13, France\\
  $^2$Yukawa Institute for Theoretical Physics,
  Kyoto University,
  Kyoto 606--8502, Japan
  \\
  $^3$Yukawa Institute for Theoretical Physics,
  Kyoto University, Kyoto 606--8502, Japan,\\
  Graduate School of Science and Technology 
    Kwansei Gakuin University, Sanda 669-1337, Japan
}

\abstract{
 We present a class of dynamical solutions for an intersecting 
D4-D8 brane system in ten-dimensional type IIA supergravity. 
The dynamical solutions reduces to a static warped 
${\rm AdS_6}\times {\rm S}^4$ geometry in a certain spacetime region.
We also consider lower-dimensional effective theories for the warped 
compactification of general $p$-brane system.
It is found that an effective $p+1$-dimensional description
is not possible in general due to the entanglement
of the transverse coordinates and the $p+1$-dimensional 
coordinates in the metric components.
Then we discuss cosmological solutions. We find a 
solution that behaves like a Kasner-type cosmological
solution at $\tau\to\infty$, while it reduces to a warped static 
solution at $\tau\to0$, where $\tau$ is the cosmic time.
}


\section{Introduction}
 \label{sec:introduction}

 Recently studies on dynamical solutions of supergravity have been
a topic of great interest.
Conventionally time dependent solutions of higher dimensional 
supergravity are discussed in the context of lower dimensional
effective theories after compactifying the internal space.
However, it is unclear how far this effective low-dimensional
description is valid. Thus it is much more desirable to discuss
the four-dimensional cosmology in terms of the dynamics of the
original higher-dimensional theory. This is particularly true in
string cosmology in which the behavior of the early universe is to be
understood in the light of string theory. Indeed, it was pointed out
that the four-dimensional effective theory for warped 
compactification of ten-dimensional type IIB supergravity 
allows solutions that cannot be obtained from solutions in 
the original higher-dimensional theories~\cite{Kodama:2005cz}. 

In the present work, we consider dynamical solutions
for intersecting D4-D8 brane systems in the ten-dimensional 
type IIA supergravity model\cite{Binetruy:2007tu}. 
In \S\,\ref{sec:Dynamical p-brane solution},
we first consider $p$-brane systems in $D$-dimensions
and derive a class of dynamical solutions under a certain metric
ansatz.
In \S\,\ref{sec:Dynamical solution for D4-D8 brane system},
focusing on intersecting D4-D8 brane systems in the ten-dimensional 
type IIA supergravity, we extend the metric ansatz used in the
previous section to intersecting branes
and obtain a class of dynamical solutions.
Then further specializing the form of the metric, 
we consider a cosmological solution.
Interestingly, this solution is found to
approach a warped static solution as $\tau\to0$
and a Kasner type anisotropic solution as $\tau\to\infty$,
where $\tau$ is the cosmic time.
Finally we conclude in \S\,\ref{sec:Conclusion}.


\section{Dynamical $\bm p$-brane solutions}
\label{sec:Dynamical p-brane solution}

We consider a gravitational theory with the metric $g_{MN}$,
dilaton $\phi$, 
and an anti-symmetric tensor field of rank $(p+2)$ in $D$
dimensions. 
This corresponds to a $p$-brane system in string theory.
The most general action for the $p$-brane system in the Einstein 
frame can be written as 
\Eq{
S=\frac{1}{2\kappa^2}\int \left(R\ast{\bf 1}_D
 -\frac{1}{2}d\phi \wedge \ast d\phi 
 -\frac{1}{2}e^{-c\phi}F_{(p+2)}\wedge\ast F_{(p+2)}\right),
\label{eq:sec2:D-dim action}
}
where $\kappa^2$ is the $D$-dimensional gravitational constant, 
$\ast$ is the Hodge dual operator in the $D$-dimensional spacetime,
and $c$ is a constant given by $c^2=4-2(p+1)(D-p-3)(D-2)^{-1}.$ 
The expectation values of fermionic fields are assumed to be zero.

To solve the field equations, we assume the $D$-dimensional metric
in the form
\Eq{
ds^2=h^a(x, y)q_{\mu\nu}dx^{\mu}dx^{\nu} 
  +h^b(x, y)u_{ij}dy^idy^j,  
 \label{eq:sec2:metric of p-brane solution}
}
where $q_{\mu\nu}$ is a $(p+1)$-dimensional metric which
depends only on the $(p+1)$-dimensional coordinates $x^{\mu}$, 
and $u_{ij}$ is the $(D-p-1)$-dimensional metric which
depends only on the $(D-p-1)$-dimensional coordinates $y^i$. 
The parameters $a$ and $b$ are given by $a=-(D-p-3)(D-2)^{-1}$, 
$b=(p+1)(D-2)^{-1}$.

Furthermore, we assume that the scalar field $\phi$ and 
the gauge field strength $F_{(p+2)}$ are given by
\Eq{
e^{\phi}=h^{-c/2},~~~~F_{(p+2)}=\sqrt{-q}d(h^{-1})
\wedge\,dx^0\wedge dx^1\wedge \cdots \wedge dx^p.
  \label{sec2:ansatz for fields:Eq}
}
Here, $q$ is the determinant of the metric $q_{\mu\nu}$.
Let us first consider the Einstein equations. 
Using the assumptions \eqref{eq:sec2:metric of p-brane solution}
and \eqref{sec2:ansatz for fields:Eq},
the Einstein equations are given by
\Eq{
hR_{\mu\nu}(\Xsp)-D_{\mu}D_{\nu} h -\frac{a}{2} 
q_{\mu\nu}\left(\triangle_{\Xsp}h + h^{-1}\triangle_{\Ysp} h\right)=0,~~
R_{ij}(\Ysp)-\frac{b}{2} u_{ij}\left(\triangle_{\Xsp} h
 +h^{-1}\triangle_{\Ysp}h \right)=0,~~
\pd_{\mu}\pd_i h=0,
 \label{sec2:components of p-brane Einstein equations:Eq}
}
where $D_{\mu}$ is the covariant derivative with respective to 
the metric $q_{\mu\nu}$, 
$\triangle_{\Xsp}$ and $\triangle_{\Ysp}$ are 
the Laplace operators on the space of 
${\rm \Xsp}$ and the space ${\rm \Ysp}$, and 
$R_{\mu\nu}(\Xsp)$ and $R_{ij}(\Ysp)$ are the Ricci tensors
of the metrics $q_{\mu\nu}$ and $u_{ij}$, respectively.
From the third equation of 
\eqref{sec2:components of p-brane Einstein equations:Eq}, 
the warp factor $h$ must be in the form
$h(x, y)= h_0(x)+h_1(y)$.
Let us next consider the gauge field.
Under the assumption \eqref{sec2:ansatz for fields:Eq}, 
we find $dF_{(p+2)}=0.$ 
Thus, the Bianchi identity is automatically satisfied.
Also the equation of motion for the gauge field becomes 
$d\left[e^{-c\phi}\ast F_{(p+2)} \right]=0.$
Hence, the gauge field equation is automatically
satisfied under the assumption~\eqref{sec2:ansatz for fields:Eq}.

Let us consider the scalar field equation.
Substituting the forms of the scalar and the gauge field 
\eqref{sec2:ansatz for fields:Eq}, 
and the warp factor $h(x, y)= h_0(x)+h_1(y)$ 
into the equation of motion for the scalar field, we obtain
\Eq{
ch^{-b}\left(\triangle_{\Xsp}h_0
+h^{-1}\triangle_{\Ysp}h_1\right)=0.
  \label{eq:sec2:p-brane scalar field equation2}
}
Thus, unless the parameter $c$ is zero, the warp factor $h$ should
satisfy the equations $\triangle_{\Xsp}h_0=0$ and $\triangle_{\Ysp}h_1=0.$
If $F_{(p+2)}\ne 0$, the function $h_1$ is non-trivial.
In this case, 
the Einstein equations reduce to  
\Eq{
R_{\mu\nu}(\Xsp)=0,~~~~R_{ij}(\Ysp)=0,~~~~D_{\mu}D_{\nu}h_0=0.
   \label{sec2:solution of p-brane Einstein equations:Eq} 
 }
On the other hand, if $F_{(p+2)}=0$, the function $h_1$ becomes
trivial. Namely the internal space is no longer
warped~\cite{Kodama:2005cz}.

Here we mention an important fact about the nature of the
dynamical solutions described in the above. In general,
we regard the $(p+1)$-dimensional spacetime to contain
our four-dimensional universe while the remaining space is
assumed to be compact and sufficiently small in size. Then 
one would usually think that an effective $(p+1)$-dimensional 
description of the theory should be possible at low energies. 
However, solutions of the field equations have the 
property that they are genuinely $D$-dimensional in the
sense that one can never neglect the dependence on $\Ysp$,
say of $h$. 
This is clear from an inspection of
Eqs.~(\ref{sec2:components of p-brane Einstein equations:Eq}).
In particular, the second equation involves the Laplacian
of $h$ with respect to the space $\Xsp$. Hence the equations
determining the internal space $\Ysp$ cannot be determined
independently from the geometry of the space $\Xsp$. 
The origin of this property is due to the existence of
a non-trivial gauge field strength which forces the
function $h$ to be a linear combination of a function
of $x^\mu$ and a function of $y^i$, instead of a product
of these two types of functions as conventionally assumed.
This fact is in sharp contrast with the case when one
is allowed to integrate out the internal space to
obtain an effective lower dimensional theory.

Finally we comment on the exceptional case of $c=0$,
which happens when $(D,p)=(10,3)$, $(11,5)$, $(11,2)$.
The scalar field becomes constant because of
the ansatz \eqref{sec2:ansatz for fields:Eq}, and 
the scalar field equation is automatically satisfied. 
Then, the Einstein equations become 
\Eq{
R_{\mu\nu}(\Xsp)=0,~~~~
R_{ij}(\Ysp)=\frac{b}{2}(p+1)\lambda u_{ij}(\Ysp),~~~~
D_{\mu}D_{\nu}h_0=\lambda q_{\mu\nu}(\Xsp),
   \label{sec2:solution of p-brane Einstein equations c=0:Eq}
 }
where $\lambda$ is a constant.
As seen from these equations, the internal space $\Ysp$ is not
necessarily Ricci flat, and the function $h_0$ becomes more complicated.
For example, when the metric $q_{\mu\nu}$ is Minkowski, $h_0$
is no longer linear in the coordinates $x^\mu$ but quadratic
in them \cite{Kodama:2005fz}.


\section{Dynamical solutions for D4-D8 brane system}
\label{sec:Dynamical solution for D4-D8 brane system}

Now we consider dynamical solutions for the D4-D8 brane system
which appears in the ten-dimensional type IIA supergravity. 
The bosonic action of D4-D8 brane system in the 
Einstein frame is given by 
\Eq{
S=\frac{1}{2{\kappa}^2}\int \left(R\ast{\bf 1}
 -\frac{1}{2}d\phi \wedge \ast d\phi 
 -\frac{1}{2\cdot 4!}e^{\phi/2}F_{(4)}\wedge\ast F_{(4)}
 -\frac{1}{2}e^{5\phi/2}m^2 \ast{\bf 1}\right).
   \label{eq:sec3:action of D4/D8}
   }
In the following, we look for a solution whose spacetime metric 
has the form
\Eq{
ds^2=h^{1/12}(z)\left[h_4^{-3/8}(x,r,z)q_{\mu\nu}dx^{\mu}dx^{\nu}
       +h_4^{5/8}(x,r,z)\left(dr^2+r^2 u_{ij}dy^idy^j+dz^2\right)\right],
   \label{sec3:D4D8 metric:Eq}
}
where $q_{\mu\nu}$ is the five-dimensional metric 
depending only on the coordinates $x^\mu$ of $\X5sp$, and 
$u_{ij}$ is the three-dimensional metric depending only 
on the coordinates $y^i$ of $\Y3sp$. 
As for the scalar field and the 4-form field strength,
we adopt the following assumptions
\Eq{
e^{\phi}=h^{-5/6}h_4^{-1/4},~~~
F_{(4)}=e^{-\phi/2}\ast \left[\sqrt{-q}d(h_4^{-1})\wedge\,dx^0\wedge dx^1\wedge dx^2 \wedge dx^3\wedge dx^4\right].
  \label{sec3:D4D8 ansatz of field:Eq}
}

Let us first consider the Einstein equations. 
Under the assumptions \eqref{sec3:D4D8 metric:Eq}
and \eqref{sec3:D4D8 ansatz of field:Eq}, the Einstein equations give 
$h_4(x,y,z)=H_0(x) + H_1(r, z).$ 
Let us next consider the gauge field $F_{(4)}$.
Under the assumptions \eqref{sec3:D4D8 metric:Eq} and 
\eqref{sec3:D4D8 ansatz of field:Eq}, the Bianchi identity 
$dF_{(4)}=0$ gives 
\Eq{
\pd_r^2 h_4+(3/r)\pd_rh_4+\pd_z^2h_4
+(1/3)\pd_z\ln h\,\pd_zh_4=0\,,~~
\pd_{\mu}\pd_rh_4=0\,,~~
\pd_{\mu}\pd_zh_4=0\,.
  \label{sec3:D4D8 Bianchi identity:Eq}
}
The last two equations are consistent with the result 
$h_4(x,y,z)=H_0(x) + H_1(r, z).$
Then the first equation \eqref{sec3:D4D8 Bianchi identity:Eq} becomes
\Eq{
\pd_r^2 H_1+(3/r)\pd_rH_1+\pd_z^2H_1
+(1/3)\pd_z\ln h\,\pd_zH_1=0\,.
  \label{eq:sec3:D4/D8 Bianchi identity2}
}
The gauge field equation $d(e^{\phi/2}\ast F_{(4)})=0$
is automatically satisfied under the 
assumption~\eqref{sec3:D4D8 ansatz of field:Eq} 
and the form of $h_4$ given by $h_4(x,y,z)=H_0(x) + H_1(r, z).$

Next we consider the scalar field equation. 
Substituting the assumptions for the metric 
\eqref{sec3:D4D8 metric:Eq}, the
scalar and gauge fields~\eqref{sec3:D4D8 ansatz of field:Eq}, 
and the form of $h_4(x,y,z)=H_0(x) + H_1(r, z)$ into 
the scalar field equation, we find
\Eq{
\triangle_{\X5sp}H_0
+(5/4)\left[(4/9)(\pd_z\ln h)^2
+(2/3)h^{-1}\pd_z^2h-m^2h^{-2}
\right]=0\,,
   \label{eq:sec3:reduced field equations1}
   }
where $\triangle_{\X5sp}$ is the Laplace operator on the space 
$\X5sp$, and we used the equation~\eqref{eq:sec3:D4/D8 Bianchi identity2}.

Inserting Eqs.~\eqref{eq:sec3:D4/D8 Bianchi identity2}
 and (\ref{eq:sec3:reduced field equations1}) into the 
Einstein equations, we find for non-trivial $H_1$,
\Eq{
R_{\mu\nu}(\X5sp)=0,~~
R_{ij}(\Y3sp)=2 u_{ij},
~~D_{\mu}D_{\nu}H_0=0\,,~~
\triangle_{\X5sp}H_0=0\,,~~
4\left(\pd_z h\right)^2/9-m^2=0\,,
~~ \pd_z^2h=0\,,
 \label{sec3:components of Einstein equations3:Eq}
}
where $R_{\mu\nu}(\X5sp)$, $R_{ij}(\Y3sp)$ are the Ricci tensors 
of the metric $q_{\mu\nu}$ and $u_{ij}$, respectively, 
$D_{\mu}$ is the covariant derivative with respective to 
the metric $q_{\mu\nu}$.
The last two equations of 
\eqref{sec3:components of Einstein equations3:Eq} 
is immediately solved to give $h(z)=3m(z-z_0)/2$, 
where $z_0$ is an integration constant
(corresponding to the position of the D8-brane).
Below we set $z_0=0$ without loss of generality.
Then \eqref{eq:sec3:D4/D8 Bianchi identity2} gives the
solution $H_1(r, z)=c_1(r^2+z^2)^{-5/3}+c_2$,
where $c_1$ and $c_2$ are constant parameters.

Let us investigate the geometrical properties of 
the D4-D8 brane system. 
As a particular solution to the 3-dimensional
metric $u_{ij}$ which satisfies
the second equation of \eqref{sec3:components of Einstein equations3:Eq}, 
we take the space $\Y3sp$ to be a three-dimensional sphere ${\rm S}^3$.
Then if we make a change of coordinates, 
$z=\tilde{r}\sin\alpha$, $r=\tilde{r}\cos\alpha$ ($0\le\alpha\le\pi/2$),
the metric reads
\Eq{
ds^2=h^{1/12}\left[h_4^{-3/8}q_{\mu\nu}dx^{\mu}dx^{\nu}
+h_4^{5/8}(d\tilde{r}^2+\tilde{r}^2d\Omega_4^2)\right],
   \label{eq:sec3:metric of near horizon limit1}
}
where $d\Omega_4^2=d\alpha^2+\cos^2\alpha d\Omega_3^2\,,$ 
$h_4(x, \tilde{r})=H_0(x)+c_1 \tilde{r}^{-10/3}\,,$ 
$h(\tilde{r}, \alpha)=(3m/2)\tilde{r}\sin\alpha\,.$ 
Here $d\Omega_3^2$ and $d\Omega_4^2$ denote the line elements 
of the three-dimensional sphere ${\rm S}^3$
and the four-dimensional sphere ${\rm S}^4$, respectively. 

Now we further define a new coordinate $U$ by $\tilde{r}^2=U^3$.
In the case $q_{\mu\nu}$ is the five-dimensional 
Minkowski metric $\eta_{\mu\nu}$, the ten-dimensional 
metric in the limit $U\rightarrow 0$ reduced to a 
warped ${\rm AdS}_6\times {\rm S}^4$ space \cite{Binetruy:2007tu}.

Let us consider the case $q_{\mu\nu}=\eta_{\mu\nu}$ in more detail.
In this case, the solution for the warp factors $h_4$ and $h$ can be obtained
explicitly as $h_4(t, \tilde{r})=\beta t+H_1(\tilde{r})$, 
$h(\tilde{r}, \alpha)=(3m/2)\tilde{r}\sin\alpha\,,$
where $H_1(\tilde{r})=c_1 \tilde{r}^{-10/3}$, and 
$\beta$ is a constant parameter. 

If we introduce a new time coordinate $\tau$ by 
$\tau/\tau_0=(\beta t)^{13/16}$, $\beta\tau_0=16/13$, 
the ten-dimensional metric is given by
\Eqr{ 
ds^2&=&h^{1/12}
\left(1+({\tau}/{\tau_0})^{-16/13}H_1\right)^{-3/8}
\left[
\left(-d\tau^2+({\tau}/{\tau_0})^{-6/13}
\delta_{ab}dx^adx^b\right)
\right.\nn\\
&&\left.
+\left(1+({\tau}/{\tau_0})^{-16/13}H_1\right)
({\tau}/{\tau_0})^{10/13}
\left(d\tilde{r}^2+\tilde{r}^2d\Omega_4^2\right)\right]\,,
 \label{eq:sec3:special metric of D4/D8-brane solution conformal time}
 }
where the metric $\delta_{ab}$ is the spatial part of the five-dimensional
Minkowski metric $\eta_{\mu\nu}$.
If we set $H_1=0$, the scale factor of the four-dimensional space
is proportional to $\tau^{-6/13}$, while that for the remaining
five-dimensional space is proportional to $\tau^{10/13}$.
Thus in the limit when the terms with $H_1$ are negligible, 
which is realized in the limit $\tau\to\infty$, we have
a cosmological solution. Although this cosmological solution is 
by no means realistic, it is interesting to note that this
cosmological solution is asymptotically static in the past
$\tau\to0$.

\section{Conclusion}
  \label{sec:Conclusion}
In this work, we investigated dynamical solutions 
of higher-dimensional supergravity models. 
We found a class of time-dependent solutions for an
intersecting D4-D8 brane system. 
These solutions were obtained by replacing a
constant $A$ in the warp factor $h=A+h_1(y)$ of
a supersymmetric solution by a function $h_0(x)$ of the 
 coordinates $x^\mu$~\cite{Kodama:2005fz},
where the coordinates $y^i$ would describe the 
internal space and $x^\mu$ would describe our universe
if the spatial dimensions of our universe were four
instead of three.
In the D4-D8 brane solution, the geometry was found to 
approach a warped static ${\rm AdS}_6 \times {\rm S}^4$
in a certain region of the spacetime.

In particular, we found an interesting solution
which is warped and static as $\tau\to0$ but
approaches a Kasner-type solution as $\tau\to\infty$,
where $\tau$ is the cosmic time.
Although the solution itself is by no means realistic,
its interesting behavior suggests a possibility that 
the universe was originally in a static state of warped 
compactification and began to evolve toward a universe
with a Kaluza-Klein compactified internal space.

Conventionally one would expect an effective theory description
in lower dimensions to be valid at low energies.
However, as clearly the case of the cosmological solution
mentioned above, the solutions we found have 
the property that they are genuinely $D$-dimensional in the
sense that one can never neglect the dependence on $y^i$,
say of $h$. 
Thus our result indicates that we have to be careful 
when we use a four-dimensional effective theory to analyse 
the moduli stabilisation problem and the cosmological 
problems in the framework of warped compactification of 
supergravity or M-theory.

\end{document}